\documentclass[aps,pra,superscriptaddress,showpacs,twocolumn]{revtex4-1}
\usepackage{amssymb}
\usepackage{graphicx}
\usepackage{dcolumn}
\usepackage{bm}
\usepackage{amsmath}
\usepackage{braket}

\usepackage[usenames]{color}
\usepackage{textcomp}
\usepackage{float}

\def\be{\begin{equation}}
	\def\ee{\end{equation}}
\def\bea{\begin{eqnarray}}
	\def\eea{\end{eqnarray}}

\usepackage[colorlinks,linkcolor=blue,anchorcolor=blue,citecolor=blue,urlcolor=blue,driverfallback=dvipdfm]{hyperref}

\begin{document}
	
\title{Orbital phases of $p$-band ultracold fermions in the frustrated triangular lattice}

\author{Jiaqi Wu}
\thanks{These authors contributed equally to this work.}
\affiliation{Department of Physics, National University of Defense Technology, Changsha 410073, P. R. China}
\author{Hui Tan}
\thanks{These authors contributed equally to this work.}
\affiliation{Department of Physics, National University of Defense Technology, Changsha 410073, P. R. China}
\author{Rui Cao}
\affiliation{Department of Physics, National University of Defense Technology, Changsha 410073, P. R. China}
\author{Jianmin Yuan}
\affiliation{Institute of Atomic and Molecular Physics, Jilin University, Changchun 130012, P. R. China}
\affiliation{Department of Physics, National University of Defense Technology, Changsha 410073, P. R. China}
\author{Yongqiang Li}
\email{li\_yq@nudt.edu.cn}
\affiliation{Department of Physics, National University of Defense Technology, Changsha 410073, P. R. China}
\affiliation{Hunan Key Laboratory of Extreme Matter and Applications, National University of Defense Technology, Changsha 410073, P. R. China}

\date{\today}
	
\begin{abstract}
Orbital degrees of freedom play an important role for understanding the emergence of unconventional quantum phases. Ultracold atomic gases in optical lattices provide a wonderful platform to simulate orbital physics. In this work, we consider spinless fermionic atoms loaded into $p$-orbital bands of a two-dimensional frustrated triangular lattice. The system can be described by an extended Fermi-Hubbard model, which is numerically solved by using the orbital version of real-space dynamical mean-field theory. Low-temperature phase diagrams are obtained, which contain stripe-, ferro- and para-orbital ordered quantum phases, due to the interplay of anisotropic hoppings and geometrical frustration. In order to understand the underlying mechanics of competing orbital orders, we derive an effective orbital-exchange model, which yields consistent explanation with our main numerical results.
\end{abstract}
	
	\maketitle

\section{introduction}
A challenging issue in condensed matter physics is to understand the behavior of strongly correlated electrons in frustrated materials. Electrons in real materials possess not only  internal spin but also orbital degrees of freedom. These two degrees of freedom are coupled to each other and related to the crystal field. On the one hand, these intricate physical effects have given rise to a diverse world, leading to a variety of attractive physical phenomena, such as unconventional superconductivity, topological insulators, colossal magnetoresistance and so on~\cite{Arg_ello_Luengo_2022,PhysRevLett.126.103201}. On the other hand, the amalgamation of these complexities renders the behavior of electrons exceedingly intricate to comprehend, posing a formidable challenge for physicists attempting to understand the behavior of strongly correlated electrons~\cite{PhysRevLett.92.216402,PhysRevLett.118.177002}.
		
Ultracold atoms provide a novel avenue to explore novel quantum physics~\cite{RevModPhys.80.885,RevModPhys.80.1215,Lewenstein_2007}. When combined with optical lattices, ultracold atomic gases serve as an unique platform for simulating and understanding many-body physics ranging from weak to strong interactions. By implementing higher Bloch bands as orbital degrees of freedom, ultracold atoms can be utilized to simulate versatile orbital physics of strongly interacting systems~\cite{2016_Li_Liu_RPP,Lewenstein2011,Dutta_2015}. 
In the past few years, extensive researches have been conducted to prepare ultracold bosonic atoms in the second Bloch band of optical lattices~\cite{PhysRevLett.99.200405,Wirth2011,PhysRevLett.114.115301,p-band_honecomb,p-topological_2021,PhysRevLett.131.226001}. Especially, Bose-Einstein condensations with resulting exotic orbital orders have been observed by combining multi-orbital setting and the complex lattice structure, such as in the honeycomb and triangular lattices~\cite{p-band_honecomb,p-topological_2021,PhysRevLett.131.226001}. However, loading ultracold fermionic atoms in the higher energy bands is challenging. Only very recently, remarkable breakthrough is achieved by successfully transferring degenerate fermions to the excited Bloch bands of a checkerboard square lattice with a lifetime up to the order of seconds~\cite{PhysRevLett.127.033201,kiefer2023ultracold}.
		
Motivated by the recent experimental progresses~\cite{PhysRevLett.127.033201,kiefer2023ultracold}, our study focuses on orbital ordering of spinless fermions loaded into the $p$-orbital bands of a two-dimensional (2D) frustrated triangular lattice. The triangular lattice, known for its frustration properties, has attracted significant interest for spinor~\cite{PhysRevB.105.205110,Xu2023,morera2024itinerant,PhysRevB.103.235132,Yang_2021,PhysRevX.12.021064,PhysRevX.10.021042,PhysRevB.103.235132,PhysRevB.91.245125,Yamamoto_2014} and orbital systems~\cite{PhysRevLett.97.190406,PhysRevLett.100.160403,PhysRevLett.100.200406}. Here, we aim to implement a comprehensive numerical understanding of the phenomena ranging from weak to strong interactions, especially the interplay of anisotropic hoppings in the parallel and the perpendicular directions in the geometrically frustrated lattice. Another open issue is whether distinct behaviors of fermions emerge, compared to bosons which manifest an orbital-skyrmion state in the identical lattice~\cite{PhysRevResearch.5.L042042}.

For a sufficiently deep lattice, the system can be effectively described by an extended multi-orbital Fermi-Hubbard model~\cite{Li_2016}. To explore the many-body physics in a frustrated triangular lattice, dynamical mean-field theory (DMFT) is developed for the $p$-orbital fermionic system. Within DMFT, local quantum fluctuations are fully taken into account to resolve competing long-range orbital orders. Actually, the non-perturbative treatment of the single-site DMFT has already been proven to be a suitable approach for spin degrees of freedom in the triangular lattice~\cite{PhysRevB.73.235107,PhysRevB.71.134422,PhysRevX.11.041013}. To tackle the translational-symmetry-breaking phases, a real-space generalization of DMFT (R-DMFT) is implemented and applied within the full range from weak to strong coupling. Our calculations of the $p$-orbital frustrated system support various quantum many-body phases, including Mott phases with stripe-, canted stripe-, and ferro-orbital orders, and a metallic phase with para-orbital ordering. Finally, we derive an orbital-exchange model in the deep Mott regime to elucidate the underlying mechanisms of competing orders.

The paper is organized as follows. In Sec.~\ref{Model and method}, we introduce the extended Fermi-Hubbard model and give an overview of R-DMFT method. In Sec.~\ref{results}, a detailed discussion of many-body properties of the system and the effective orbital-exchange model are presented. Finally, we make a summary in Sec.~\ref{Summary}.

		\begin{figure}[t]
			
	\includegraphics[width=\columnwidth]{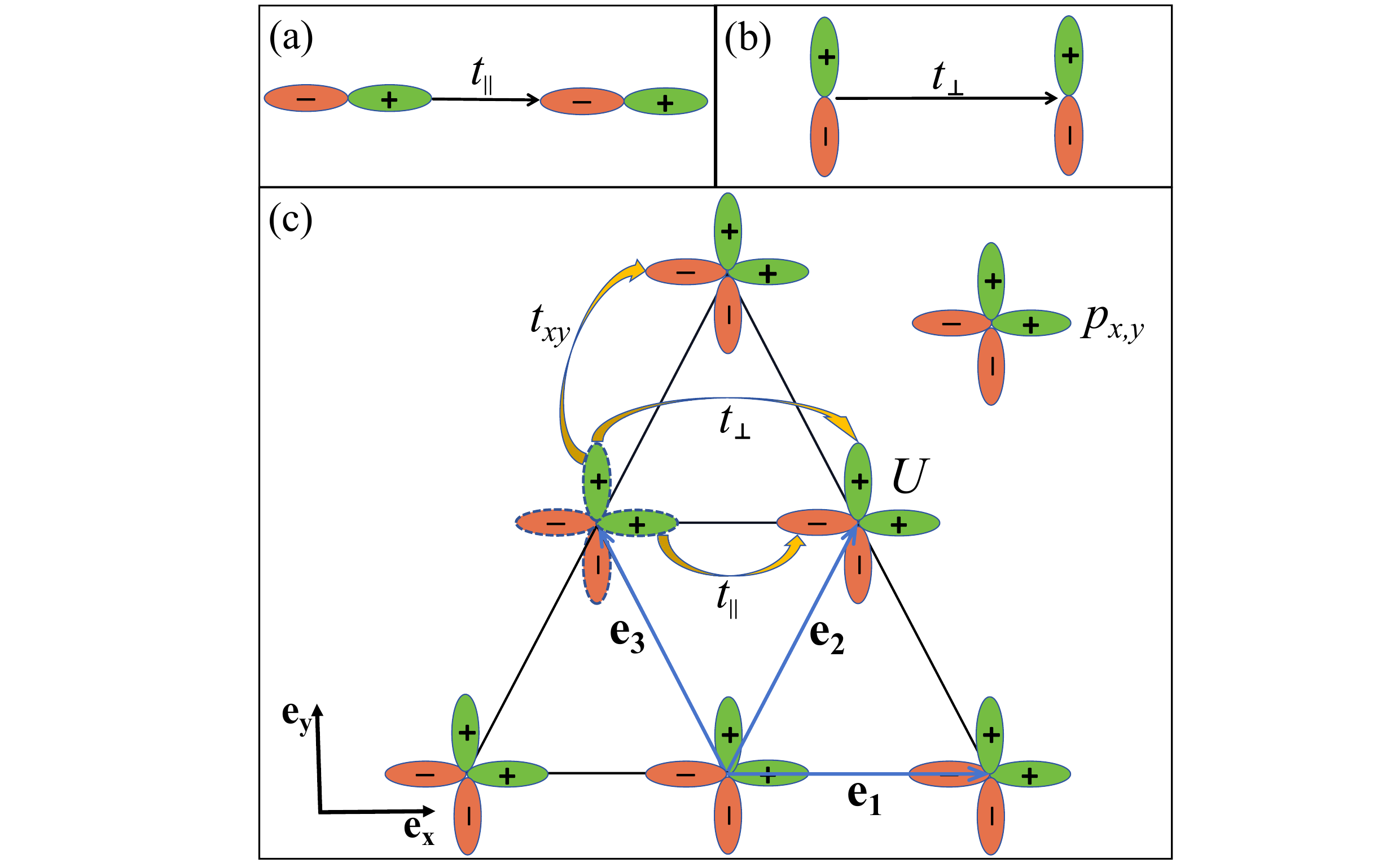}\caption{\label{FIG1}Sketch of the $p$-orbital fermions loaded into the two-dimensional triangular lattice. (a)(b) Two types of hopping matrix elements of $p$-orbital fermions in the parallel (a) and perpendicular directions (b), respectively. (c) The system possesses {\it normal} hopping terms $t_{\parallel,\perp}$, orbital-flipping processes $t_{xy}$, and onsite interaction $U$ for the $p_x$ and $p_y$ orbitals.}
	\end{figure}		
	\section{Model and method}
          \label{Model and method}
	\subsection{MODEL}
           \label{Model}
We consider a single-component fermionic gas loaded into $p$-orbital bands of a triangular lattice, as shown in Fig.~\ref{FIG1}. Here, a strong confinement is added to freeze the motional degree of freedom in the third direction, realizing a 2D triangular-lattice system. In contrast to the internal spin degrees of freedom, orbital physics is characterized by orbital degeneracy and spatial anisotropy which cause the difference between hopping amplitudes along different directions. For a sufficiently deep lattice, the system can be described by an extended multi-orbital Fermi-Hubbard model~\cite{Li_2016}
	\begin{eqnarray}
			\label{eq:Ham}
			\hat H &=&t_{\parallel} \sum_{m, {\bf r} }  \hat p_{m,{\bf r}} ^\dag \hat p_{m,{\bf r} + {\bf e}_m}- t_{\perp}\sum_{m, {\bf r}}  \hat p^{^\prime\dagger}_{m, {\bf r}}  \hat p^\prime_{m,{\bf r} + {\bf e}_m}
			+ {\rm H.c.}  \nonumber \\
			&+& {U}  \sum_{\bf r} \hat n_{x,{\bf r}} \hat n_{y,{\bf r}} - \sum_{\sigma,\bf r} \mu_{\sigma} \hat n_{\sigma ,\bf r}.		
	\end{eqnarray}
Here $\sigma$ = $\{x,y\}$, and the unit vectors $ \mathbf{e} _{1}= \mathbf{e} _{x}$, $\mathbf{e}_{2}=\frac{1}{2}\mathbf{e} _{x} + \frac{\sqrt{3}}{2} \mathbf{e} _{y}$, and $ \mathbf{e} _{3}=-\frac{1}{2}\mathbf{e} _{x} + \frac{\sqrt{3}}{2}\mathbf{e} _{y}$, where the lattice constant is set as the unit of length. The lattice annihilation operators $\hat p_{m,{\bf r}} \equiv (\hat p_{{x},{\bf r}}{\bf e}_x  + \hat p_{{y},{\bf r}}{\bf e}_y)\cdot {\bf e}_m$, and $\hat p^\prime_{m,{\bf r}} \equiv ( \hat p_{{x},{\bf r}}{\bf e}_x  + \hat p_{y,{\bf r}}{\bf e}_y\big)\cdot {\bf e}^\prime_m$ with ${\bf e}^\prime_1={\bf e}_y$ and ${\bf e}^\prime_{2,3}=-\frac{\sqrt{3}}{2}{\bf e}_x \pm \frac{1}{2} {\bf e}_y$. $t_{\parallel}$ and $t_{\perp}$ denote the $p$-orbital hopping amplitudes of the nearest-neighbor couplings along the parallel and the perpendicular directions, respectively. Note here that the $p$-orbital model includes orbital-flipping hopping processes between the $p_x$ and $p_y$ orbitals along the ${\bf e}_2$ and ${\bf e}_3$ bonds~\cite{p-band_honecomb}, denoted as $t_{xy}\equiv\frac{\sqrt{3}}{4}(t_{\perp}+t_{\parallel})$, which is in contrast to the $s$-band spinor fermions in a triangular lattice. $\hat p_{\sigma,\mathbf{r}}$ ($\hat p^{\dagger}_{\sigma,\mathbf{r}}$) is the annihilation (creation) operator for the $\sigma$ orbital at site $\mathbf{r}$, $\hat n_{\sigma,\mathbf{r}}$ the number operator, and $\mu_\sigma $ the chemical potential. $U$ is the local Hubbard interaction for atoms in $p_{x}$ and $p_{y}$ orbitals, which is contributed by the $p$-wave scattering due to Pauli exclusion principle~\cite{PhysRevLett.100.160403,PhysRevLett.100.200406}. We remake here that $p$-wave interaction is typically weak for ultracold atoms, and its stability against $p$-wave Feshbach resonances is limited by three-body losses in the strongly interacting region~\cite{PhysRevLett.90.053201}. However, $p$-wave interactions can still be tuned in a wide regime via Feshbach resonance, and unitary $p$-wave interactions between spinless fermions are recently achieved in a multi-orbital optical lattice~\cite{Venu_2023}.

	\subsection{METHOD}
   \label{METHOD}
To understand this extended Fermi-Hubbard model, dynamic mean-field theory is developed to calculate many-body ground states of the $p$-orbital system, described by Eq.~(\ref{eq:Ham}). DMFT is a exact theory in infinite dimension and a good approximation for finite dimension. The key point of DMFT is to map the many-body lattice system to a single-site impurity connected to non-interacting fermionic baths, and then the impurity problem is solved self-consistently. DMFT takes fully account of local quantum fluctuations of the strongly correlated system, but neglects nonlocal spatial fluctuations. Though DMFT neglects nonlocal fluctuations, it has been proven to be a good approximation for the triangular lattice due to its big coordination number $z=6$~\cite{PhysRevB.73.235107,PhysRevB.71.134422,PhysRevX.11.041013}. In order to investigate various exotic orbital ordered phases which break lattice-translational symmetry, we implement a real-space version of DMFT~\cite{Snoek_2008,PhysRevLett.100.056403,PhysRevLett.105.065301,PhysRevLett.121.093401,PhysRevA.105.063308,
PhysRevA.101.063611,Han2023,PhysRevA.108.033314}. Within R-DMFT, the self-energy is a local quantity but position-dependent. Local physical quantities can be obtained after solving the single-site impurity problem, and the physics of the impurity site is given by the local effective action $S_{eff}^{\bf 0}$. In order to derive the local effective action $S_{eff}^{\bf 0}$ for the impurity site {$\bf {0}$}, one needs to integrate out the remaining lattices' degrees of freedom ($\bf r\neq 0$) in the partition function
\begin{eqnarray}
			\label{eq:Seff}
			\frac{1}{Z_{eff}}e^{-S_{eff}^{\bf 0}}\equiv\frac{1}{Z}\int\prod_{\bf r\neq 0,\sigma}D{p}_{\sigma,\bf r}^{\star}D{p}_{\sigma,\bf r}e^{-S},
\end{eqnarray}
where $S$ is the action of the full system, and ${p}_{\sigma,\bf r}^{\star},{p}_{\sigma,\bf r}$ are Grassmann variables describing fermions. $S_{eff}^{\bf 0}$ can be obtained from the standard derivation~\cite{1996Dynamical}. Here, a brief presentation is shown, where the effective action of the system can be written as

\begin{widetext}
	\begin{equation} \label{ action of the system}
		\begin{aligned}
		S[{\bf p}^{\star},\bf p] &=\int_{0}^{\beta}\,d{\tau}  \sum_{\sigma,\bf r}{{p_{\sigma,{\bf r}} ^\star ( \partial_{\tau} -\mu_{\sigma})  p_{\sigma,\bf r}}}  + \sum_{\sigma_1,\sigma_2, {\bf r},{\bf e_{i}} } t_{\bf e_{i}}^{\sigma_1,\sigma_2} {( p_{\sigma_1,{\bf r}} ^\star  p_{\sigma_2,{\bf r} + {\bf e}_i} + p_{\sigma_2,{\bf r} + {\bf e}_i}^\star p_{\sigma_1,{\bf r}} )}\\
&+{U}\sum_{\bf r}  n_{x,{\bf r}}  n_{y,{\bf r}}.
		\end{aligned}
	\end{equation}
For brevity of formula derivation, we represent various hopping amplitudes of atoms in different directions as $t_{\bf e_{i}}^{\sigma_1,\sigma_2}$, which contains spin-flipping and conserving terms. After intergating out the lattices' degrees of freedom, the effective action for the impurity site could be derived, which is given by
\begin{equation} \label{effective action for the impurity}
	\begin{aligned}
	{S_{eff}^{\bf 0}=\int_0^{\beta}d\tau_1 d\tau_2\sum_{\sigma_1,\sigma_1^{\prime}}\left(\begin{array}{c}p^{\star}_{\sigma_1,{\bf 0}}(\tau_1)\\p_{{\sigma_1,{\bf 0}}}(\tau_1)\end{array}\right)^T	 \mathcal{G}_{{\bf 0},\sigma_1,\sigma_1^{\prime}}^{-1}(\tau_1-\tau_2)\left(\begin{array}{c}p_{\sigma_1^{\prime},{\bf 0}}(\tau_2)\\p_{\sigma_1^{\prime},{\bf 0}}^{\star}(\tau_2)\end{array}\right)+{U}n_{x,{\bf 0}}  n_{y,{\bf 0}},}
	\end{aligned}
\end{equation}
\end{widetext}
where $\quad$ $\quad$ $\quad$ $\quad$ $\quad$ $\quad$ $\quad$ $\quad$ $\quad$ $\quad$ $\quad$ $\quad$ $\quad$ $\quad$ $\quad$ $\quad$ $\quad$ $\quad$ $\quad$
\begin{widetext}
\begin{equation}
	\begin{aligned} \label{local non-interacting propagator}
		&\mathcal{G}_{{\bf 0},\sigma_1,\sigma_1^{\prime}}^{-1}(\tau_1-\tau_2)=\\
		&\left(\begin{array}{ccc}
			(\partial_{\tau_2}-\mu_{\sigma_1})\delta_{\sigma_1,\sigma_1^{\prime}}\delta(\tau_1-\tau_2)+\sum_{\sigma_2,\bf e_i,\sigma_2^{\prime},\bf e_i^{\prime} }{G}_{{\bf e_i }\sigma_2,{\bf e_i^{\prime} }\sigma_2^{\prime}}^{\bf 0}(\tau_1-\tau_2) & 0\\
			\\
			0 & \sum_{\sigma_2,\bf e_i,\sigma_2^{\prime},\bf e_i^{\prime} }{G}_{{\bf e_i^{\prime} }\sigma_2^{\prime},{\bf e_i }\sigma_2}^{\bf 0}(\tau_2-\tau_1)\\
		\end{array}\right) , \nonumber
	\end{aligned}
\end{equation}
with ${G}_{{\bf e_i }\sigma_2,{\bf e_i^{\prime} }\sigma_2^{\prime}}^{\bf 0}(\tau_1-\tau_2)=-\langle \hat T\hat{p}_{\sigma_2,{\bf e_i }}(\tau_1)\hat{p}_{\sigma_2^{\prime},\bf e_i^{\prime}}^\dagger(\tau_2)\rangle_{\bf 0}$ describing the Green's functions for the $p$-orbital fermions, and $\left\langle ...\right\rangle_{\bf 0}$ the expectation value in the cavity system without the impurity site.
\end{widetext}
Here, $ \mathcal{G}_{{\bf 0},\sigma_1,\sigma_1^{\prime}}^{-1}(\tau_1-\tau_2)$ is a local non-interacting propagator interpreted as a dynamical Weiss mean field which simulates the effects of all other sites. $S_{eff}^{\bf 0}$ allow us to calculate all the local correlation functions of the original Hubbard model. For the reason that it is difficult to resolve this effective action analytically in the practice, we utilize the effective mean-field hamiltonian and map the original Hubbard model onto a set of single-impurity Anderson models~\cite{PhysRevB.80.245109,PhysRevB.84.144411}
\begin{widetext}
\begin{equation} \label{Anderson impurity Hamiltonian}
	\begin{aligned}
		{\hat{H}_A}^{(\iota)}&=
		{U_f} \hat n_x \hat n_y -\sum_{\sigma} \mu_{\sigma} \hat n_{\sigma} +\sum_{l,\sigma} \epsilon_{l,\sigma}^{(\iota)} \hat a_{l,\sigma}^{\dagger} \hat a_{l,\sigma} + \sum_{l,\sigma} \left( V_{l,\sigma}^{(\iota)} \hat a_{l,\sigma}^{\dagger}  \hat c_{\sigma}  + W_{l,\sigma}^{(\iota)}  \hat a_{l,\bar{\sigma}}^{\dagger}  \hat c_{\sigma}  +\text { H.c. } \right)
	\end{aligned}
\end{equation}
\end{widetext}
for each site $\iota$. The notation $\sigma=\{x,y\}$ represents one of the two $p$-band components, and $l=\{1,2,3\cdots n_s\}$ denotes bath index for each $p$-orbital component. The noninteracting fermions in the bath are described by operators $\hat a_{l,\sigma}$ with energies $\epsilon_{l,\sigma}$, and local impurity fermions are described by operators $\hat c_{l,\sigma}$. $V_{l,\sigma}$ and $W_{l,\sigma}$ describe the spin-conserving and flipping couplings between the baths and the impurity site, respectively. The impurity Hamiltonian can be solved self-consistently using exact diagonalization (ED) as a solver. By diagonalizing the Anderson Hamiltonian in the Fock basis, the corresponding solution of the impurity model can be obtained. Here the truncation of bath sites $n_s=4$ is chosen mainly in our calculations. After diagonalization, the local Green's functions in the Lehmann representation can been obtained
\begin{equation}	
\begin{aligned}
&G_{ A}^{(\iota),\sigma \sigma^{\prime}}(i \omega_n)= \\
&-\frac{1}{Z} \sum_{m,n} \frac {\left\langle m \right| \hat c_{\sigma} \left| n \right\rangle  \left\langle n \right| \hat c_{\sigma^{\prime}}^{\dagger} \left| m \right\rangle}{E_n - E_m -i\hbar \omega_n} (e^{-\beta E_n }+e^{-\beta E_m }), \\
\end{aligned}
\end{equation}
where $Z$ is the partition function, $\omega_n=(2n+1)\pi /\beta$ denotes fermionic Matsubara frequency, and $\beta$ is the inverse temperature.  
Then the local self-energy for each site can be obtained via the Dyson equation
\begin{equation}
\Sigma_A^{(\iota)} (i \omega_n)= \mathcal{G}_A^{(\iota)-1}(i \omega_n) -G_A^{(\iota)-1}(i \omega_n),
\end{equation}
where $\mathcal{G}_A^{(\iota)}(i \omega_n)$ denotes the non-interacting Weiss Green's function of the Anderson impurity site. The Weiss Green's function can be expressed as
\begin{subequations}
	\begin{align}
		&\mathcal{G}_A^{(\iota)-1\sigma\sigma}(i\omega_n) =i\omega_n+\mu_\sigma+\sum_{l}\frac{|V_{l,\sigma}^{(\iota)}|^2}{i\omega_n-\varepsilon^{(\iota)}_{l,\sigma}} + \frac{|W_{l,\sigma}^{(\iota)}|^2}{i\omega_n-\varepsilon_{l,\bar{\sigma}}^{(\iota)}}, \\
		&\mathcal{G}_A^{(\iota)-1 \sigma\bar{\sigma}}(i\omega_n) =\sum_{l}\frac{V_{l,\bar{\sigma}}W^\ast_{l,\sigma}}{i\omega_n-\varepsilon^{(\iota)}_{l,\bar{\sigma}}} + \frac{W_{l,\bar{\sigma}}V^\ast_{l,\sigma}}{i\omega_n-\varepsilon^{(\iota)}_{l,\sigma}}.
	\end{align}
\end{subequations}
Here, $\sigma$ = $\{x,y\}$, $\bar x$ = $y$, and $\bar y$ = $x$. In the framework of R-DMFT, we assume that the impurity self-energy $\Sigma_A^{(\iota)} (i \omega_n)$  coincides with lattice self-energy $\Sigma_{lattice}^{(\iota)}(i \omega_n)$. Specifically, after we obtain the self-energy $\Sigma_{lattice}^{(\iota)}(i \omega_n)$ for each site, we collect them in the real-space self-energy matrix $\mathbf{\Sigma}_{lattice} (i \omega_n)$. Next, we can employ the Dyson equation in the real-space representation to
compute the interacting lattice Green's function
\begin{equation}
\mathbf{G}_{lattice}^{-1}(i \omega_n)= \mathbf{G}_{0}^{-1}(i \omega_n) - \mathbf{\Sigma}_{lattice} (i \omega_n),
\end{equation}
where the non-interacting lattice Green's function $\mathbf{G}_{0}^{-1}(i \omega_n)=( i \omega_n \sigma_z + \bm{\mu})-\mathbf{t}$,  with the matrix of hopping $\mathbf{t}$ determined by the lattice structures. Note here that the boldface quantities denote matrices with site-dependent elements. The self-consistency R-DMFT loop is closed by the Dyson equation to obtain a new local non-interacting propagator.
New Anderson impurity parameters are then updated by minimizing the difference between old and new propagators, and the procedure is then iterated until convergence is reached.

\section{results}
\label{results}

\subsection{Mott-metal transition}
\label{Mott-metal transition}
In the first part, we study  Mott-metal transition of the $p$-band fermionic atoms in a 2D triangular lattice ranging from weak to strong interactions, based on R-DMFT. To distinguish different quantum phases, we compute the following quantities, including the double occupancy  \\
	\begin{eqnarray}
			\label{eq:double occupancy}
			{D_{\mathbf{occ}}} &=&\frac{1}{N_{\rm lat}}\sum_{\bf r}\left<{\hat n_{x,\mathbf{r}}\hat n_{y,\mathbf{r}}}\right>,		
	\end{eqnarray}
and the quasi-particle weight
\begin{eqnarray}
			\label{eq:quasiparticle weight}
			Z_{\mathbf{r}}=\frac1{1-\frac{\partial\Sigma_{\mathbf{r}}(\omega)}{\partial\omega}}=\frac1{1-\frac{\Im m\Sigma_{\mathbf{r}}(i\omega_0)}{\omega_0}}.		
	\end{eqnarray}
Here, $\left<...\right>$ denotes the ensemble average, and $N_{\rm lat}$ is the number of lattice sites. In addition, the one-particle spectral function $\rho (\omega)$ can also be implemented in the R-DMFT calculations to quantify the localization of the many-body system. The one-particle spectral function can be determined by retarded Green's function, which can be calculated by mapping the imaging frequency Matsubara Green's function to the real frequency,
\begin{eqnarray}
			\label{eq:dos}
			\rho(\omega)&=&-\frac{1}{\pi}Tr Im{\bf G}^R(\omega + i\delta),	
	\end{eqnarray}
with $\delta$ denoting the inverse lifetime of the quasi-particle. In our work, we mainly focus on the half-filled case with balanced mixtures, i.e., $\mu_\sigma = \mu$ for each lattice site. To resolve orbital orders of the quantum many-body phases, we focus on low-temperature condition, and set $\beta$$t_{\parallel}=25$. The stability of orbital order is verified against temperatures, and no quantitative difference is found for even lower temperatures. The largest lattice size $N_{\rm lat}=24\times24$ is chosen in our calculations.

\begin{figure}[t]
	\includegraphics[width=\columnwidth]{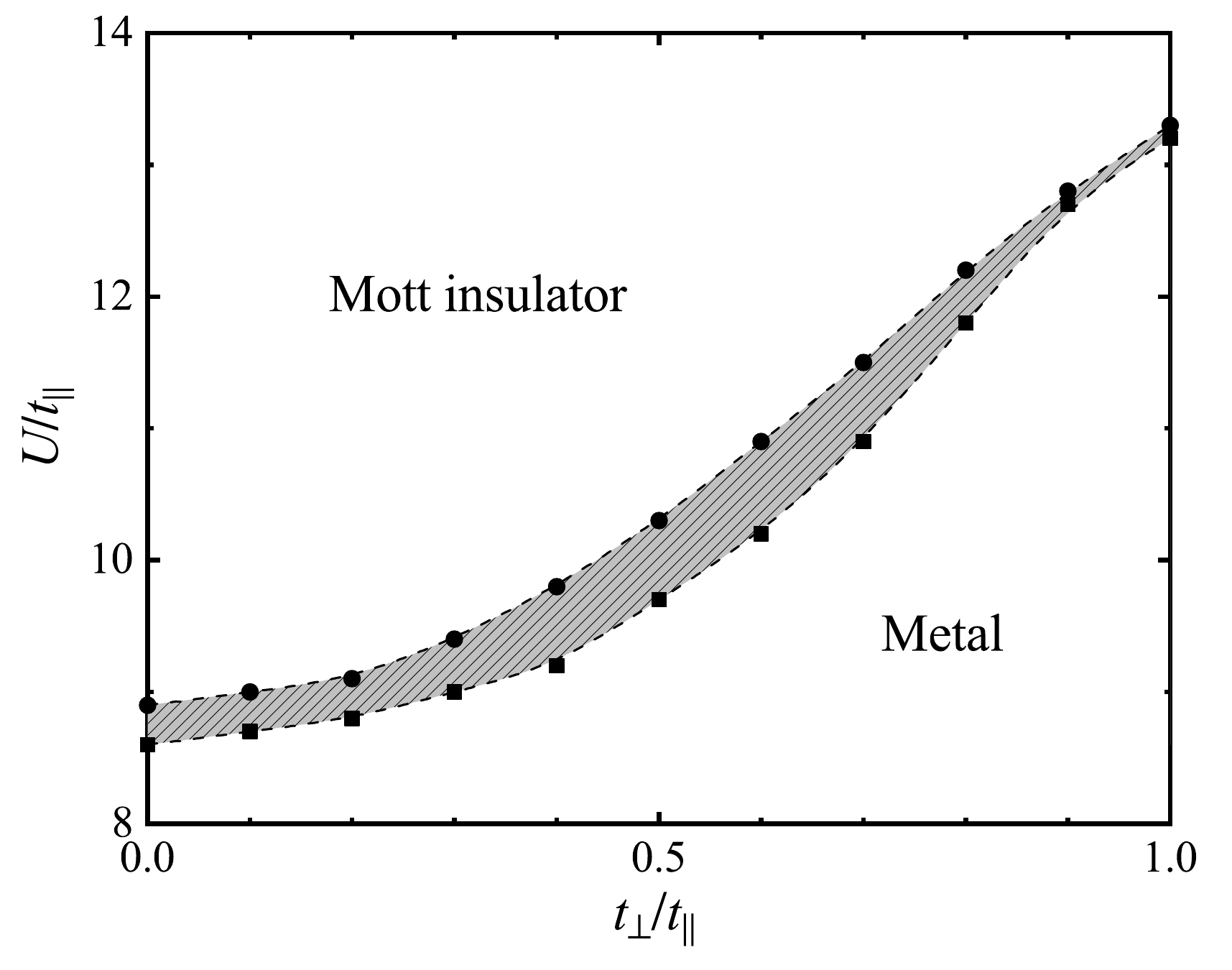}
	\caption{Paramagnetic quantum phase diagram of the $p$-orbital fermions in a 2D triangular lattice as a function of Hubbard interactions $U/t_{\parallel}$ and hopping amplitudes $t_{\perp}/t_{\parallel}$. The shaded region is the metal-insulator coexistence region depending the initial conditions, where the upper and the lower boundaries correspond to the solutions starting from the metallic and deep Mott phases, respectively. We choose the temperature $\beta$$t_{\parallel}=25$.}
\label{FIG2}
\end{figure}

Fig.~\ref{FIG2} displays the many-body phase diagram of $p$-orbital fermions in terms of interaction $U/t_{\parallel}$ and hopping $t_{\perp}/t_{\parallel}$ for filling $n\equiv\frac{1}{N_{\rm lat}} \sum_{{\sigma,\bf r}} \langle {\hat n}_{\sigma,\bf r}\rangle=1$. To describe the Mott-metal transition, we use {\it paramagnetic} solution within R-DMFT, which prohibits spontaneous symmetry breaking~\cite{PhysRevB.105.235102,PhysRevB.103.125132}.  In this phase diagram, we observe two distinct quantum phases, i.e., Mott-insulating and metallic phases, characterized by double occupancy $D_{\textbf {occ}}$, quasi-particle weight $Z_{\mathbf{r}}$, and one-particle spectrum function $\rho(\omega)$. As expected, the system favors the metallic phase for weak interactions, and the Mott phase develops in the strongly interacting regime. In addition, we observe a coexistence region with both Mott and metallic states being stable with R-DMFT, whose solution depends on the initial conditions of the self-consistency loop.

\begin{figure}[t]
	\includegraphics[width=\columnwidth]{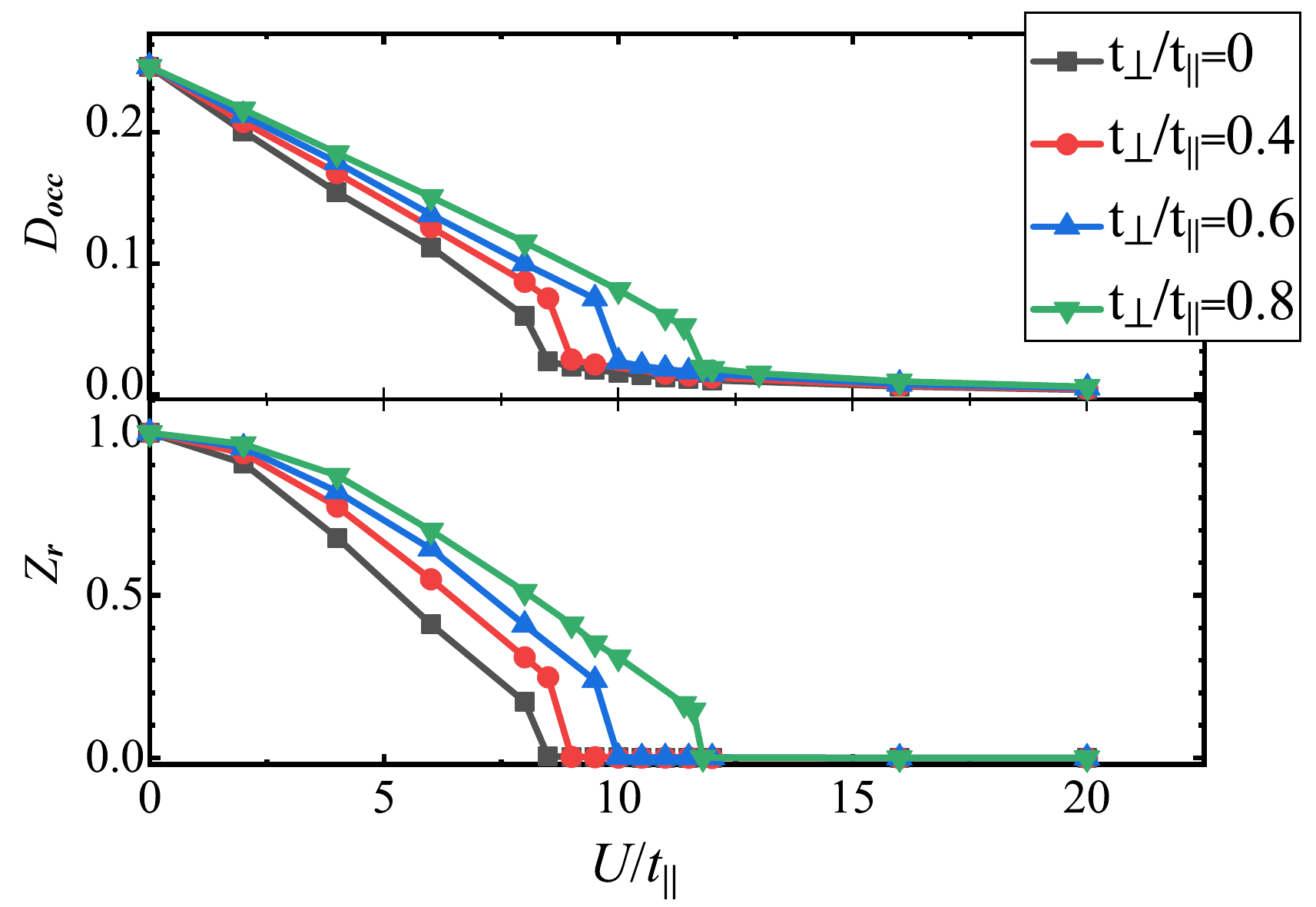}
	\caption{Double occupancy $D_\textbf{occ}$ (upper panel) and quasi-particle weight $Z_\textbf{r}$ (lower panel) as a function of Hubbard interaction $U/t_{\parallel}$ for different hopping amplitudes $t_{\perp}/t_{\parallel}$. We observe a first-order transition from the metallic to the Mott phase.}
\label{FIG3}
\end{figure}

The coexistence region indicates that the Mott-metal transition is first order. Fig.~\ref{FIG3} shows double occupancy and quasi-particle weight as a function of interaction $U$ for different hopping amplitudes $t_{\perp}/t_{\parallel}$. When the system is in the noninteracting limit with $U/t_\parallel=0$, two fermionic components are decoupled, so double occupation $D_{\mathbf{occ}}= 1/4$ and quasi-particle weight $Z_{\mathbf{r}}= 1$, which is a signature of completely itinerant nature for the half-filled fermions in the optical lattice~\cite{PhysRevB.100.125141}. $D_{\textbf{occ}}$ and $Z_{\mathbf{r}}$ drop rapidly when the interaction increases, as a result of the energy cost for the doubly occupied atoms, signifying increased localization of the system. For larger values of $U/t_{\parallel}$, the R-DMFT self-energy possesses the characteristic negative divergent low-frequency behavior, so $Z_{\mathbf{r}}$ approaches zero nearly. As a result, we can identify a cusp in $D_{\mathbf{occ}}$ and $Z_{\mathbf{r}}$ as a function of interaction, which corresponds to a phase transition from a metallic to a Mott phase. The discontinuous change of the observations indicates it is a first-order phase transition.

\begin{figure}[t]
	\includegraphics[width=\columnwidth]{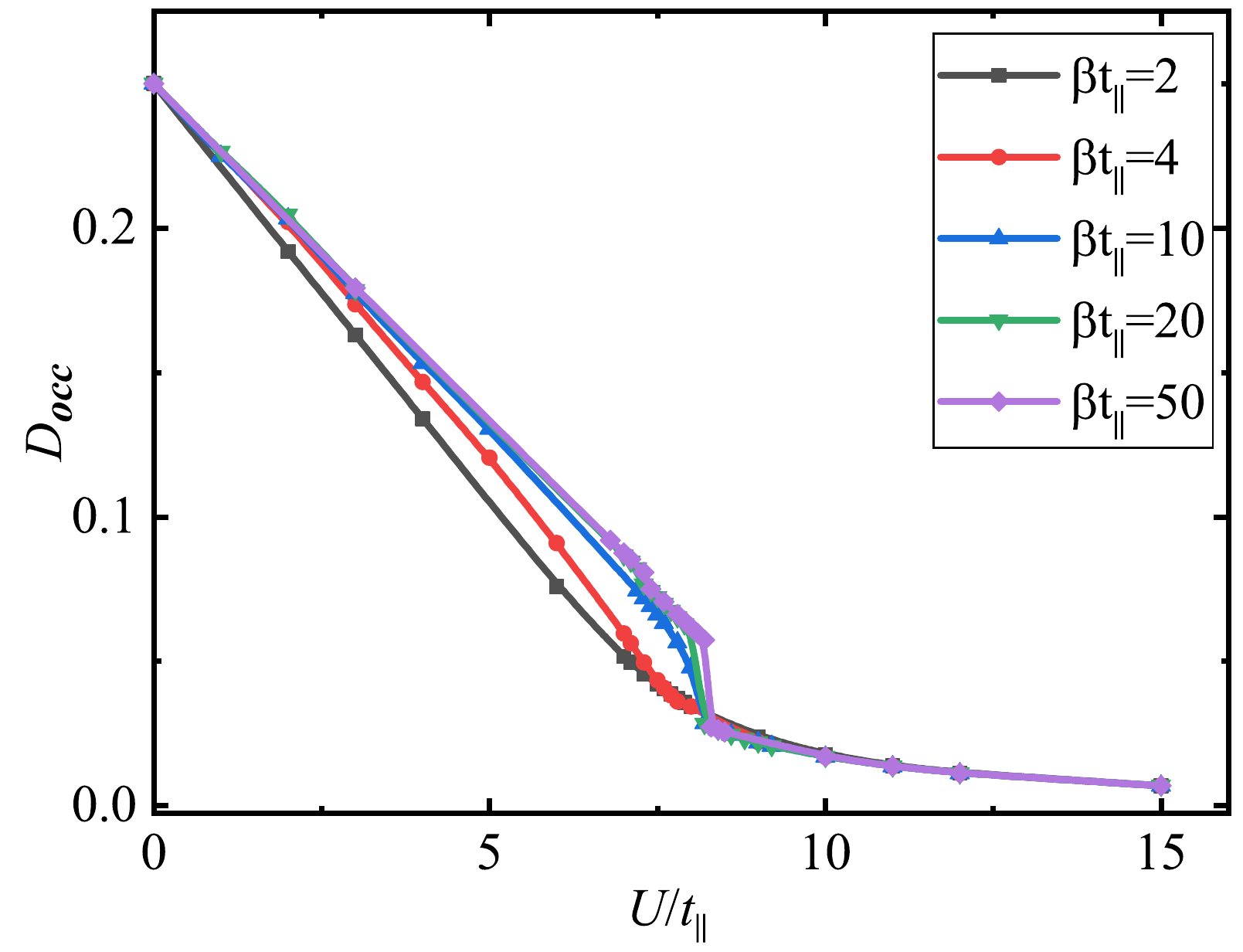}
	\caption{Double occupancy $D_\textbf{occ}$ as a function of the Hubbard interaction $U/t_{\parallel}$ for different temperatures at a fixed hopping amplitude $t_{\perp}/t_{\parallel}$=0. Upon increasing the temperature, a crossover from the metallic to the Mott phase occurs instead of the phase transition.}
\label{FIG4}
\end{figure}
Next, we study the influence of temperature on the phase transition. Generally, a metal-Mott transition of the orbital system occurs for low temperatures. When the temperature is high, it is expected that a crossover occurs instead of the phase transition~\cite{Hofstetter_2018}. Fig.~\ref{FIG4} shows the double occupancy of $p$-orbital fermions with $t_\perp/t_\parallel=0$ for different temperatures. For temperatures $\beta$$t_{\parallel} = 2$ or $\beta$$t_{\parallel} = 4$, we observe the phase transition changes to a smooth curve, indicating the Mott-metal crossover arising at higher temperatures. As the temperature drops, the curve exhibits a discontinuous behavior, and we can identify a cusp which is treated as the onset of a Mott transition instead of a crossover. We observe that different curves collapse to a single one when the interaction is larger than the critical value $U/t_\parallel\approx8.3$, which indicates that the temperatures considered here do not affect the double occupancy distinctly at large interactions. Note here that the evolution of the double occupancy with temperature is in line with the results of $s$-band spinful fermions loaded into the 2D triangular lattice~\cite{PhysRevA.82.043625}.

\begin{figure}[t] 	
	\includegraphics[width=\columnwidth]{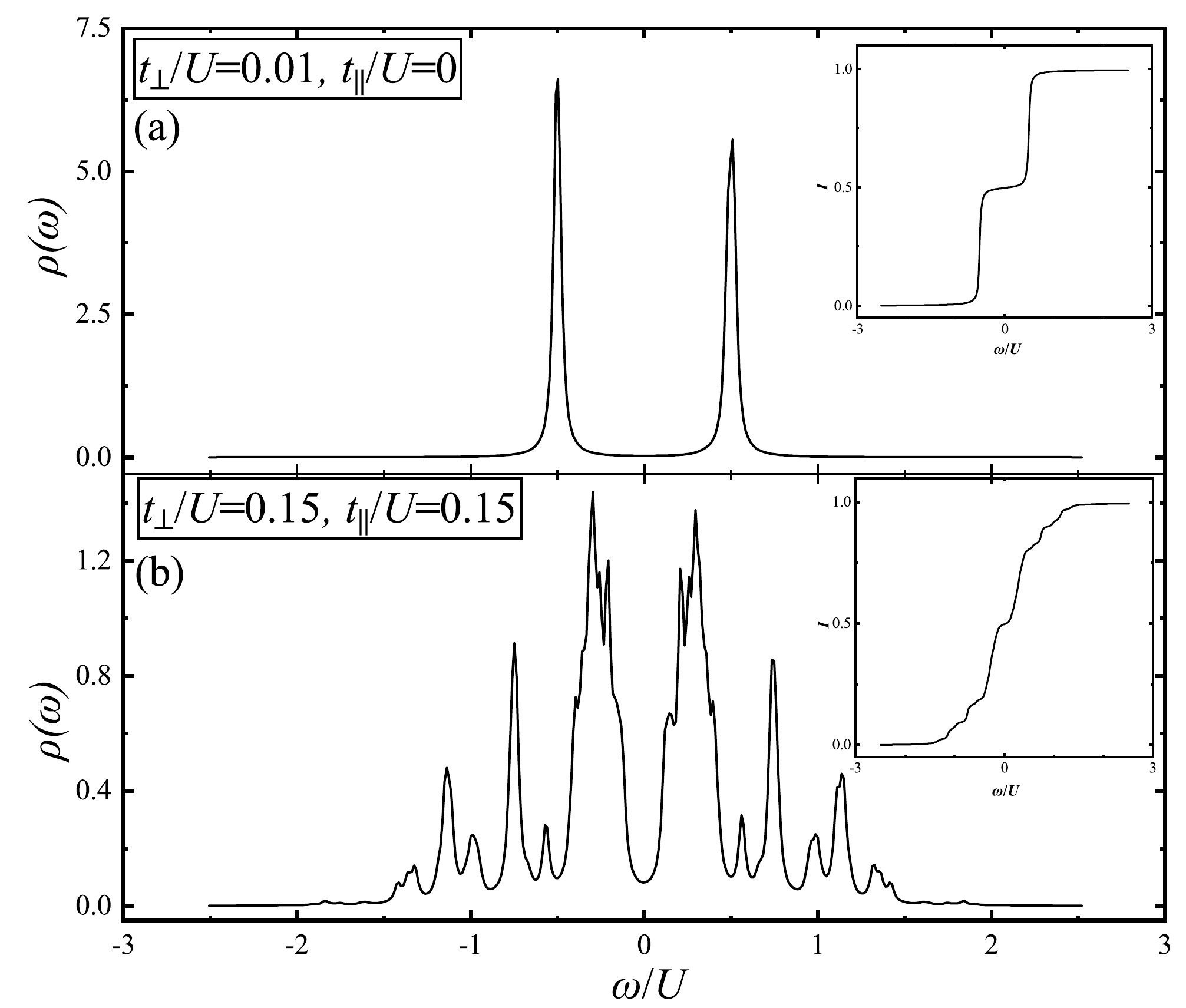}
	\caption{One-particle spectrum as a function of frequency $\omega$ for the Mott-insulating (a) and the metallic phases (b), where the parameters are $t_{\parallel}/U=0.01$ and $t_{\perp}/U=0$ (a), and $t_{\parallel}/U=0.15$ and $t_{\perp}/U=0.15$ (b). The insets show the corresponding integrated one-particle spectrum $I(\omega) = 1/2\int_{-\infty}^\omega\rho(\omega^{\prime})d\omega^{\prime}$ for each case.
	}
\label{FIG5}
\end{figure}

The one-particle spectral function $\rho(\omega)$ is also calculated in our work. In order to show the spectral structure explicitly, we make a change and set interaction $U$ as the energy unit. In the deep Mott-insulating regime, the spectrum of the finite-size Anderson model consists of two peaks separated by the energy gap $U$, as shown in Fig.~\ref{FIG5}(a), which corresponds to the energy cost for adding a particle to the system in the Mott limit. With the decrease of the interaction, the system enters the metallic phase, as shown in Fig.~\ref{FIG5}(b). In the metallic regime, the spectrum consists of a large number of peaks, where the finite Hubbard gap vanishes. Note here that, due to the finite truncation of the bath orbitals in the ED solver, the one-particle spectrum only contains coarse-grained information about exact solution and fine details of the spectrum are poorly reproduced~\cite{1996Dynamical}. The insets of Fig.~\ref{FIG5} show the integrated one-particle spectral function $I(\omega) = 1/2
\int_{-\infty}^\omega\rho(\omega^{\prime})d\omega^{\prime}$. The integral $I(\omega)$ ends up with a stable value of 1, which is known as the spectral sum rule and indicates that one particle occupies precisely the equivalent of one quantum state.

\subsection{Orbital-ordered phase diagram}
\label{Orbital-ordered phase diagram}
\begin{figure}[t] 	
	\includegraphics[width=\columnwidth]{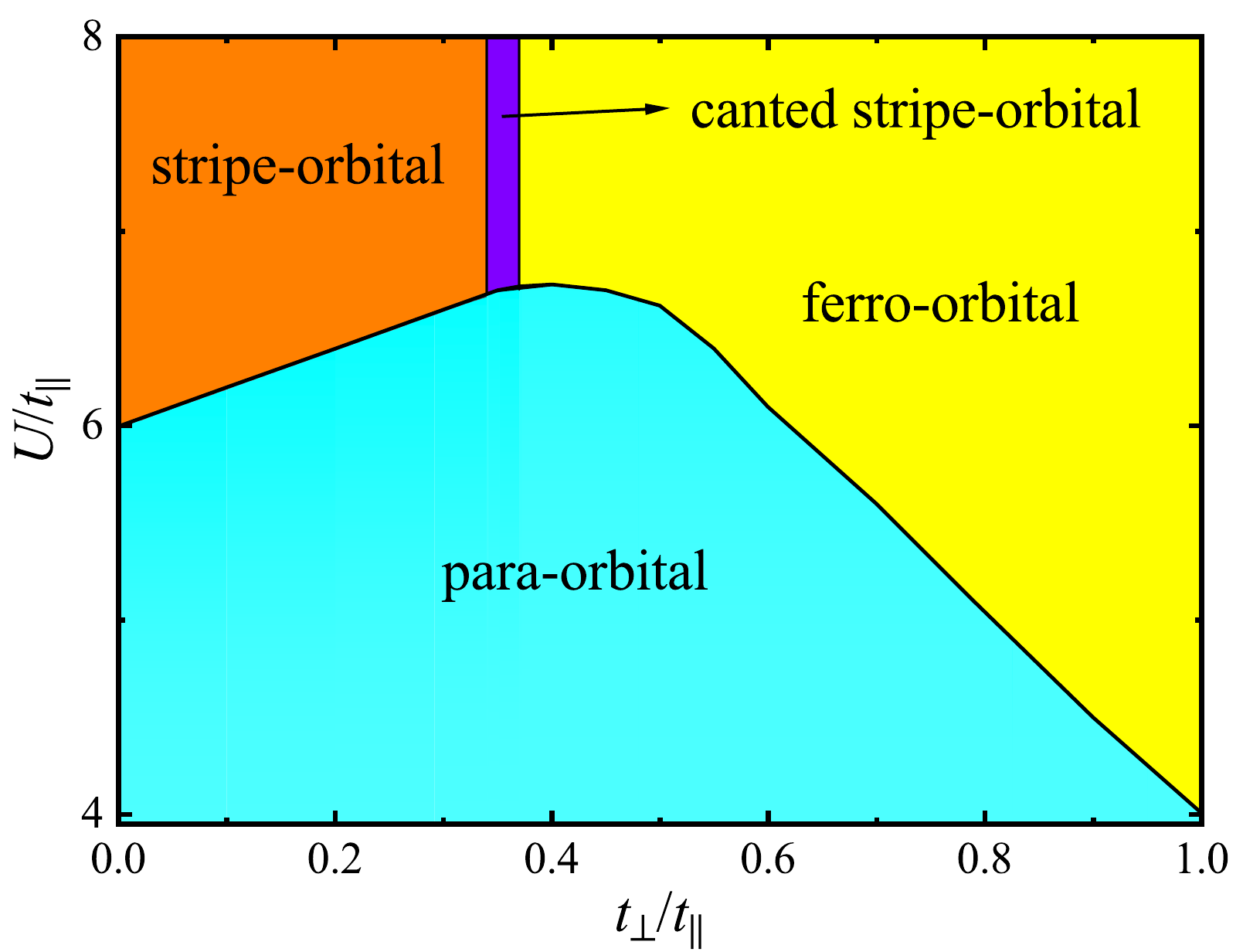}
	\caption{Phase diagram of the extended Hubbard model in the 2D triangular lattice for the half-filled case in the $t_{\perp}-U$ plane, obtained by R-DMFT, with the energies given in units of the hopping parameter $t_{\parallel}$. The system prefers para-orbital metallic phase, and various Mott phases with stripe-, canted stripe-, and ferro-orbital orders. We choose the temperature $\beta$$t_{\parallel}=25$.}
\label{FIG6}
\end{figure}

R-DMFT is a non-perturbative method and includes local quantum fluctuations of the correlated quantum many-body system. Besides {\it paramagnetic} solutions within R-DMFT, it is expected that this method can capture orbital ordering of the strongly correlated $p$-orbital fermionic system as well. Therefore, we also calculate {\it magnetic} solutions within R-DMFT, which resolves the orbital ordered phases with spontaneous symmetry breaking~\cite{PhysRevB.105.235102,PhysRevB.103.125132}. To distinguish different long-range orders of the many-body phases, we calculate local orbital polarization $\langle \hat S_{\bf r} \rangle = \left[ \langle \hat S^X_{\bf r} \rangle, \langle  \hat S^Y_{\bf r}\rangle, \langle  \hat S^Z_{\bf r} \rangle  \right]$, where the pseudospin operators from the orbital degrees of freedom are utilized, with $ \hat S_{\bf r}^Z= \frac{1}{2}(\hat p^{\dagger}_{x,\mathbf{r}} \hat p_{x,\mathbf{r}} - \hat p^{\dagger}_{y,\mathbf{r}} \hat p_{y,\mathbf{r}})$, $ \hat S_{\bf r}^X=\frac{1}{2}(\hat p^{\dagger}_{x,\mathbf{r}} \hat p_{y,\mathbf{r}} + \hat p^{\dagger}_{y,\mathbf{r}} \hat p_{x,\mathbf{r}})$, and $ \hat S_{\bf r}^Y=\frac{1}{2i}(\hat p^{\dagger}_{x,\mathbf{r}} \hat p_{y,\mathbf{r}} - \hat p^{\dagger}_{y,\mathbf{r}} \hat p_{x,\mathbf{r}})$. Note here that $\hat S_{\bf r}^Y$ is indeed the orbital angular momentum operator. Accordingly, we also define the static orbital-order structure factor $ S_{\vec q} =\left| \frac{1}{N_{\rm lat}} \sum_{\bf r} \langle \hat S_{\bf r} \rangle e^{i \vec {q} \cdot \vec r_{\bf r}} \right| $. Fig.~\ref{FIG6} displays the orbital-ordered many-body phase diagram in terms of interaction $U/t_\parallel$ and hopping $t_{\perp}/t_\parallel$ for filling $n\equiv\frac{1}{N_{\rm lat}} \sum_{{\sigma,\bf r}} \langle {\hat n}_{\sigma,\bf r}\rangle=1$. For weak interaction, the system is in the para-orbital phase with $\langle \hat S_i \rangle = 0$, which preserves both time-reversal and lattice-translational symmetries. In the strongly interacting Mott regime, various orbital orders develop, including a collinear stripe-orbital ordered phase with $\langle \hat S^{X,Z}_{\bf r} \rangle\neq0$ by breaking lattice-translational symmetry, and a ferro-orbital ordered phase which carries the orbital angular momentum $\langle \hat S^{Y}_{\bf r} \rangle\neq0$ by breaking time-reversal symmetry. In between these two orbital ordered phases, a canted-stripe ordered phase exists in a small parameter regime with $\langle \hat S^{X,Y,Z}_{\bf r} \rangle\neq0$, which breaks both time-reversal and lattice-translational symmetries.

\begin{figure}[t]
	\includegraphics[width=\columnwidth]{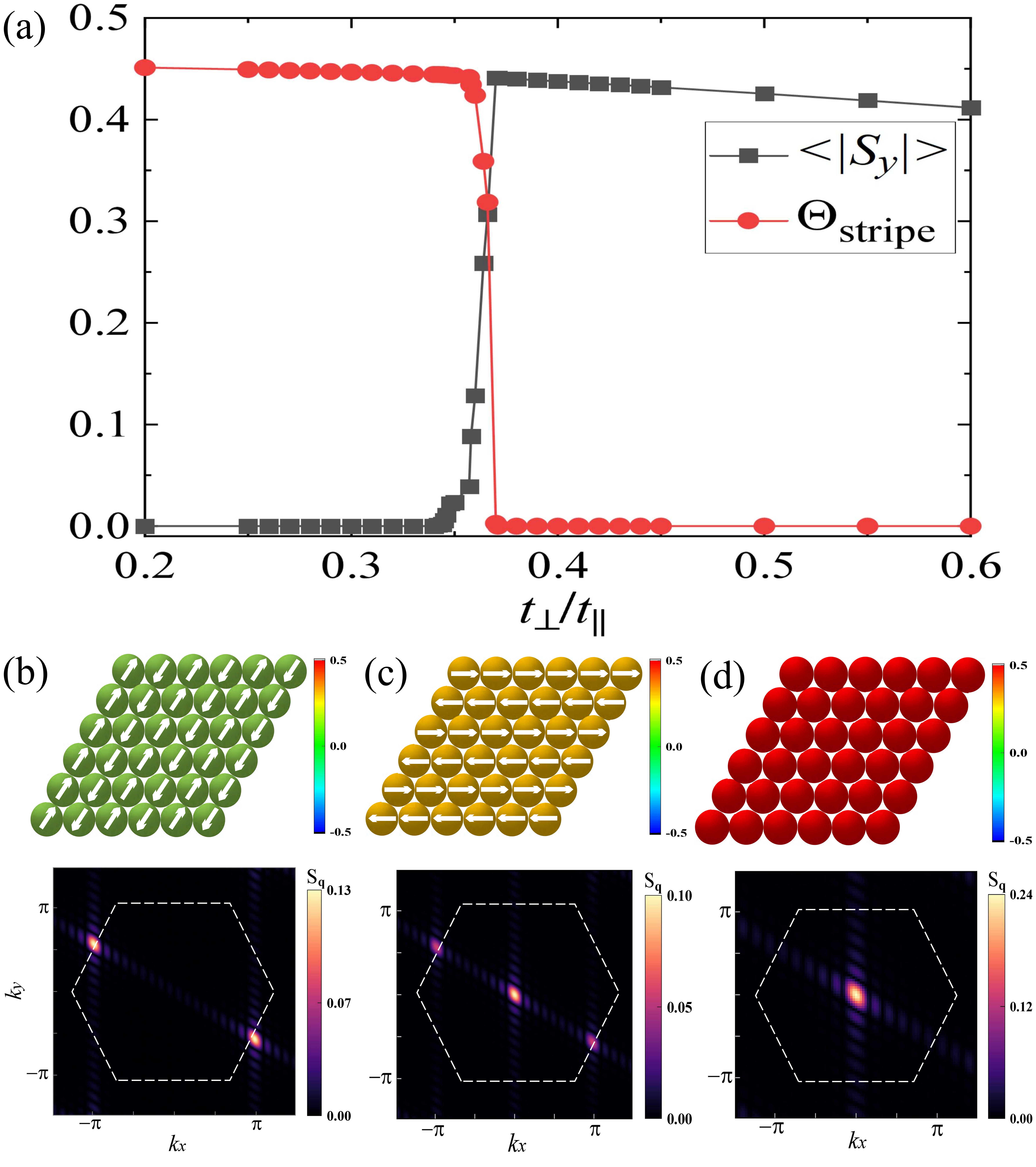}
	\caption{ (a) Stripe-to-ferro-orbital phase transition of ultracold fermionic gases in the $p$-orbital bands of a 2D triangular lattice. (b-d) real-space distributions of orbital textures (upper panels) and the corresponding static structure factor $S_{\vec q}$ (lower panels) for the stripe- (b), canted stripe- (c), and ferro-orbital (d) ordered phases, respectively. Here, the arrows represent the $xz$-component of the local orbital vector and the color denotes the $y$-component. Other parameters are $t_{\perp}/t_{\parallel}$=0.2 (b), $t_{\perp}/t_{\parallel}$=0.36 (c), and $t_{\perp}/t_{\parallel}$=0.4 (d), and $U/t_{\parallel}$=8.}
\label{FIG7}
\end{figure}

To quantify phase boundaries between the orbital ordered states, we introduce order parameters $\langle\lvert S_y\rvert\rangle=\sum_{\bf r} \lvert\langle \hat S^{Y}_{\bf r} \rangle\rvert/{N_{\rm lat}}$ and ${\Theta_{\rm stripe}}=\sum_{{\bf r},{\bf e}_m}\frac{1}{4{{{N_{\rm lat}}}}}\sqrt{(\langle \hat S^{X}_{\bf r} \rangle-\langle \hat S^{X}_{{\bf r}+{\bf e}_m} \rangle)^2+(\langle \hat S^{Z}_{\bf r} \rangle-\langle \hat S^{Z}_{{\bf r}+{\bf e}_m} \rangle)^2}$. 
Fig.~\ref{FIG7}(a) shows order parameters as a function of hopping amplitudes. We observe a stripe-to-ferro-orbital phase transition with increasing the hopping $t_\perp/t_\parallel$. The corresponding orbital ordering for different quantum phases are shown in the real [upper panel of Fig.~\ref{FIG7}(b-d)] and momentum spaces [lower panels of Fig.~\ref{FIG7}(b-d)]. The stripe phase breaks the lattice-rotational and translational symmetries [Fig.~\ref{FIG7}(b)]. We remark here that the stripe phase predicted here agrees with the previous work which is in the limit of $t_{\perp}/t_{\parallel}=0$~\cite{PhysRevLett.100.200406}. The ferro-orbital phase is characterized by a non-zero $\langle  \hat S^Y_i\rangle$, which carries the orbital angular momentum and breaks time-reversal symmetry [Fig.~\ref{FIG7}(d)]. It may provide a new perspective for realizing orbital quantum anomalous Hall effect in high-orbital optical lattices~\cite{PhysRevLett.101.186807}.

\subsection{Perturbation theory at $U \gg t_{\parallel},t_{\perp}$ }
\label{Perturbation theory}

In order to understand the orbital ordered phases in the deep Mott regime, an effective orbital-exchange model of the system in Eq.~(\ref{eq:Ham}) is derived at half filling. The Hamiltonian is divided as $\hat H = \hat H_U + \hat H_t$, where
\begin{subequations}
	\begin{align}\label{H_U}
	&\hat{H_t} = t_{\parallel} \sum_{m, {\bf r} }  \hat p_{m,{\bf r}} ^\dag \hat p_{m,{\bf r} + {\bf e}_m}- t_{\perp}\sum_{m, {\bf r}}  \hat p^{^\prime\dagger}_{m, {\bf r}}  \hat p^\prime_{m,{\bf r} + {\bf e}_m}
	+ {\rm H.c.},  \\
	&\hat{H_U} ={U}  \sum_{\bf r} \hat n_{x,{\bf r}} \hat n_{y,{\bf r}}.
	\end{align}
\end{subequations}
The effective orbital-exchange Hamiltonian is obtained by considering $\hat H_{\rm t}$ part as a perturbation to the full Hamiltonian in the strong coupling limit $t_{\perp,\parallel}\ll U$~\cite{PhysRevLett.91.090402,essle_2005,Mila2011,PhysRevLett.111.205302}. We denote the single-atom occupied operator as $\hat P$, which projects the whole Hilbert space to a subspace with only one atom per site. We call the subspace as $P$ space. Obviously, $\hat{P}$ is a hermitian projector which commutes with ${\hat H}_{\rm U}$. We denote by $\hat Q=1 - \hat P$ a complementary operator of $\hat P$, which projects the whole many-body Hilbert space to a subspace with more than one atom per site at least. We call the subspace as $Q$ space.

After a standard derivation, we obtain an effective model for the evolution of the original Hamiltonian in the subspace in which $\hat P$ projects at half filling
\begin{equation} \label{Heff}
\hat{H}_{\rm eff} =  \hat P  \hat{H}_t \hat Q \frac{1}{-\hat Q \hat{H}_U\hat Q } \sum_{n=0}^{\infty} \left( \hat Q \hat{H}_t\hat Q  \frac{1}{-\hat Q \hat{H}_U\hat Q } \right)^n \hat Q  \hat{H}_t \hat P.
\end{equation}
Keeping terms up to second order $\mathcal{O}\left( t^2 / U \right)$, we finally obtain an effective orbital-exchange model
\begin{equation} \label{effective hamiltion}
	\hat H_{\rm eff}=\underset{\langle i,j\rangle}\sum J_x\mathcal{S}^x_{i}\mathcal{S}^x_{j} + J_y\mathcal{S}^y_{i}\mathcal{S}^y_{j} +J_z\mathcal{S}^z_{i}\mathcal{S}^z_{j},
\end{equation}
where $\left\langle i,j \right\rangle$ denotes the nearest-neighbor sites $i$ and $j$, and the Heisenberg exchange coupling terms $J_{x,y,z}$ are given in the Appendix. The $J_z$ term dominates in the regime $t_\parallel\ll t_\perp$ or $t_\perp \ll t_\parallel$, where the system with $J_z > 0$ prefers a stripe-orbital ordered phase. This conclusion is consistent with previous results~\cite{PhysRevLett.100.160403,PhysRevLett.100.200406}. In the regime $t_\parallel \approx t_\perp$, the $J_{y}$ term is pronounced, where the ground state of the system with $J_{y} < 0$ prefers ferro-orbital order, reminiscent of the $XXZ$ model for a spinor system in the triangular lattice~\cite{Yamamoto_2014,Melko_2007,PhysRevB.91.081104}. The corresponding physics is consistent with our numerical simulations within R-DMFT. In contrast to the bosonic system with orbital-skyrmion state in the regime $t_\parallel\approx t_\perp$ with $J_z >0$~\cite{PhysRevResearch.5.L042042}, the Ising-type frustration occurs in the regime $t_\parallel\ll t_\perp$ or $t_\perp \ll t_\parallel$ for the fermions, forming stripe-orbital ordering in the triangular lattice.

\section{conclusion}
\label{Summary}
In this work, we explore the extended Fermi-Hubbard model with ultracold fermions loaded into the $p$-orbital bands of a two-dimensional triangular lattice. To investigate this system, a real-space version of dynamical mean-field theory is developed and applied, which enables us to study orbital ordering of the strongly correlated fermionic system. Our calculations show that the system is in the para-orbital phase at weak interaction. In the strongly correlated regime, the system develops stripe- and ferro-orbital phases, as a result of the interplay of the orbital anisotropy and geometric frustration. Interestingly, the ferro-orbital phase spontaneously breaks time-reversal symmetry, and may provide a new perspective for realizing intrinsic quantum anomalous Hall effect in high-orbital optical lattices. To better understand various orbital ordered phases, we derive the effective orbital-exchange model based in the deep Mott regime, whose conclusion is in agreement with numerical calculations. Considering the experimental realizations of $p$-orbital fermions~\cite{PhysRevLett.127.033201} and the Hubbard model on a triangular lattice~\cite{Xu2023,Yang_2021}, it is expected that the predicted orbital textures can be realized and probed by the standard Bragg spectroscopy technique in the near future~\cite{1999_Ketterle_PRL}. 

\begin{acknowledgments}
We acknowledge helpful discussions with Irakli Titvinidze, Andrii Sotnikov, Xiaopeng Li, and Wei Yi. This work is supported by the National Natural Science Foundation of China (Grants No. 12074431, and 11774428), and Excellent Youth Foundation of Hunan Scientific Committee under Grant No. 2021JJ10044. We acknowledge the ChinaHPC for providing HPC resources that have contributed to the research results reported within this paper.
\end{acknowledgments}

\newpage
\begin{widetext}
	\section{Appendix}
	\renewcommand{\theequation}{S\arabic{equation}}
	\renewcommand{\thefigure}{S\arabic{figure}}
	\renewcommand{\bibnumfmt}[1]{[#1]}
	\renewcommand{\citenumfont}[1]{#1}
	\setcounter{equation}{0}
	\setcounter{figure}{0}

\subsection{Effective orbital-exchange model}	
In this part, we derive the effective orbital-exchange model in the deep Mott regime. Eq.~{\eqref{Heff}} reads
\begin{equation} \label{second-order}
		\hat{H}_{\rm eff} =  \hat P  \hat{H}_t \hat Q \frac{1}{-\hat Q \hat{H}_U\hat Q }  \hat Q \hat{H}_t\hat Q  .
\end{equation}
For the half-filled case, we consider a two-sites problem, and then the subspace $\mathcal{H}_P$ is given by
\begin{equation} \label{H_P}
	\mathcal{H}_P:\{ \ket{p_x,p_x},\ket{p_x,p_y},\ket{p_y,p_x},\ket{p_y,p_y}\},
\end{equation}
where $\ket{p_\sigma,p_{\sigma^\prime}}$ denotes the orbital state $p_x$ or $p_y$ in two adjacent sites . The subspace $\mathcal{H}_Q$, where lattice site is doubly occupied with two atoms, is
\begin{equation} \label{H_Q}
	\begin{aligned}
	\mathcal{H}_Q:
	&\left\lbrace \ket{p_x p_y;0},\ket{0;p_x p_y}\right\rbrace.
\end{aligned}
\end{equation}

From these two subspaces, the matrix form of $\hat P  \hat{H}_t \hat Q$, $\hat Q \hat{H}_U\hat Q$ and $\hat Q \hat{H}_t\hat Q$ can be obtained. Eq.~(\ref{second-order}) yields the effective orbital-exchange model. When we consider two-neighbouring sites along the bond direction of ${\bf e}_1$, the effective Hamiltonian up to second order $\mathcal{O} \left(t_{\parallel,\perp}^2/U \right)$ is given by
\begin{eqnarray}
\label{eq:e1}
{\hat H^{\bf{e}_1}}_{\rm eff}&=&\underset{i}\sum J_x\mathcal{S}^x_{i}\mathcal{S}^x_{i+{\bf e}_1}+J_y\mathcal{S}^y_{i}\mathcal{S}^y_{i+{\bf e}_1}+J_z\mathcal{S}^z_{i}\mathcal{S}^z_{i+{\bf e}_1},
\end{eqnarray}
where the three coupling strengths respectively read
\begin{equation}
\begin{aligned}
 J_{x}=J_{y}=-\frac{4t_{\perp}t_{\parallel}}{U},J_{z}=\frac{2(t_{\perp}^2+t_{\parallel}^2)}{U}.
\end{aligned}
\end{equation}
We remark here that $J_{z}$ is antiferro-orbital exchange interaction, and $J_{x,y}$ is ferro-orbital exchange interaction. This is equivalent to the $XXZ$ model of the spinor systems in the triangular lattice.

For the other two bond directions, they can be easily obtained by rotating the coordinate of orbital polarization operators, for bond direction forming an angle of $\theta_m$ with the x axis, $\boldsymbol{\mathcal{S}}$ becomes
\begin{eqnarray}
\label{Sz}
\tilde{\mathcal{S}}^z_{\bf r}&=& {\rm sin} \left(2\theta_m \right) { \mathcal{S}}^x_{\bf r}+{\rm cos} \left(2\theta_m \right){ \mathcal{S}}^z_{\bf r} , \\
\label{Sx}
\tilde{\mathcal{S}}^x_{\bf r}&=& {\rm cos} \left(2 \theta_m \right) { \mathcal{S}}^x_{\bf r}-{\rm sin} \left(2 \theta_m \right){ \mathcal{S}}^z_{\bf r}, \\
\label{Sy}
\tilde{\mathcal{S}}^y_{\bf r}&=& { \mathcal{S}}^y_{\bf r},
\end{eqnarray}
where ${\rm \theta_m}=0, \frac{2}{3}\pi$ and $\frac{4}{3}\pi$ for the bond directions ${\bf e}_1$, ${\bf e}_2$ and ${\bf e}_3$, respectively.

\end{widetext}

\bibliography{reference}

\end{document}